# A Novel Semantic Software for Astronomical Concepts

M. Heydari-Malayeri [1], N. Moreau[1], F. Le Petit [2]

(1) LERMA, Paris Observatory, 61 Avenue de l'Observatoire, 75014 Paris, France.
(2) LUTH, Paris Observatory, 92195 Meudon cedex, France.

**Abstract**: We have created a new semantic tool called AstroConcepts, providing definitions of astronomical concepts present on Web pages. This tool is a Google Chrome plug-in that interrogates the *Etymological Dictionary of Astronomy and Astrophysics*, developed at Paris Observatory. Thanks to this tool, if one selects an astronomical concept on a web page, a pop-up window will display the definition of the available English or French terms. Another expected use of this facility could be its implementation in Virtual Observatory services.

## 1. Introduction

There is a need for a reliable and comprehensive reference source providing the definitions of all the concepts, from the oldest to the newest, used in astronomy and astrophysics. Valuable efforts have so far been made to gather a large number of astronomical concepts; for example the four-volume work by Klecze & Kleczkova (1990), which also gives the equivalents in six European languages, but does not provide the definitions of concepts. We can also mention the four-volume set *Encyclopedia of Astronomy and Astrophysics* (Murdin, 2001), which deals with many topics with ample explanations in articles aimed at astronomers. There are also numerous astronomical dictionaries intended for amateur astronomers and a larger public (among the most recent ones, one can mention, e.g., Riadpath, 2012 and Daintith & Gould 2009). Despite their valuable utility, these references, however, do not meet the requirements of completeness and up-to-dateness not to mention other considerations, including the fact that they are not interdisciplinary oriented, nor freely accessible.

An exhaustive concept base would make life easier, because instead of searching definitions through sparse subset references, one could have access to all the available definitions at one go. Besides practical use, it would be interesting to collect in a single source all the astronomical concepts that have ever been created. It would be even more helpful to be able to access such a source freely online as a database that would get enriched on a daily basis through the inclusion of new concepts created in current research work. This aspect contrasts with usual dictionaries which are updated only at intervals of several years.

Such a concept source is being developed at Paris Observatory in the form of an interactive database (MySql/Php) called *An Etymological Dictionary of Astronomy and Astrophysics* (Heydari-Malayeri, 2012). Currently it contains the definitions of about 10,000 English entries, with their French and Persian equivalents. Each definition is checked by specialist astronomers at Paris Observatory or other international research institutes. In this sense it is a collaborative international endeavor that should serve various research, educational, and development goals, such as those defined by the International Astronomical Union, in particular its Commission 46, which seeks to promote the development and improvement of astronomical education at all levels through the world (Jones, 2012). More detailed information about various aspects of the dictionary is given elsewhere (Heydari-Malayeri, 2009). Hence in Section 3, below, we only highlight some of the distinctive marks of this work.

## 2. A Browser Embedded Astrophysical Dictionary

We developed a Google Chrome extension, called AstroConcepts, giving access to definitions of the *Etymological Dictionary of Astronomy and Astrophysics* from any web page. AstroConcepts, created by one of the authors, Nicolas Moreau (2012), interrogates a SKOS (Simple Knowledge Organization System (Isaac & Summers, 2009) version of the *Etymological Dictionary of Astronomy and Astrophysics*. This formal language is recommended by the International Virtual Observatory Alliance (IVOA, Derriere et al., 2009) to create Knowledge Organization System information in a Semantic Web compatible form. SKOS is a common data model for expressing the basic structure and content of concept schemes, such as thesauri, classification schemes, taxonomies, subject heading systems, and other similar types of controlled vocabularies. SKOS is also meant to develop and re-use standard tools (concept-based search engines, browsers) for the exploitation of the Knowledge Organization Systems published on the Semantic Web. Moreover it hides the complexity of Web Ontology Language (OWL), which is a more expressive language for defining the syntax and semantics of vocabularies.

### 2.1 What is lurking behind a Web page word?

An immediate application of the AstroConcepts tool is to have access to the definition of any astronomical term, or that in a related field of knowledge, on any Web page. Its installation is a very simple operation requiring just a couple of clicks, the instructions for which are presented by Moreau (2012). Once installed and activated, readers can underline a concept on a Web page to get access to its definition as provided by the *Etymological Dictionary of Astronomy and Astrophysics*. Selected words can be either in French or English, while the definitions, appearing in a pop-up window, are in English. Put metaphorically, the tool offers a sort of radioscopy of the page words to the user.

### 2.2 Virtual Observatory

Virtual Observatory is an international initiative by the astronomical community to allow global electronic access to the available astronomical data archives of space and ground-based observatories. It also aims to enable data analysis techniques through a coordinating entity that provides common standards, wide-network bandwidth, and state-of-the-art analysis tools. The Virtual Observatory is also intended for re-using data for scientific objectives different from the original ones, in order to optimize the science return of astronomical observations. The Virtual Observatory's capabilities are enabled through the use of standard protocols for registering the existence and location of data and for requesting data that satisfies the user's interests. These standards are developed on an international basis through the IVOA. The cornerstone of the Virtual Observatory is interoperability (Egret & Genova, 2001). Interoperability is the ability of different types of computers, networks, operating systems, and software applications to work together by exchanging and sharing information in a standardized, accurate, and effective manner. AstroConcepts is an initiative in line with the requirements of interoperability.

The AstroConcepts tool has been developed thanks to the experience of VO-Paris Data Center staff in Semantic Web. Indeed, several IVOA efforts rely on a Web semantic layer. The Semantic Working Group at the IVOA as well as the VO-Theory Interest Group develop several SKOS vocabularies to define concepts. One of the aims of these efforts is to help users to discover astronomical data thanks to commonly used concepts. Eventually, the SKOS vocabulary on which AstroConcepts rely could be used in VO-Tools and astronomical services to provide definitions of astronomical concepts as well as refine the interoperability between services.

## 3. Some features of the dictionary

The AstroConcepts tool also allows the interested reader to have direct access, through a click, to the *Etymological Dictionary of Astronomy and Astrophysics* for additional information. In fact the dictionary is intended for professional and amateur astronomers, university students in astrophysics, as well as terminologists and linguists, especially those interested in the etymology of Indo-European languages. Some of the particularities of this work are specified below.

### 3.1 Word filiation

Indeed, the origin, history, and the way in which a term is composed provide the reader with an additional dimension of the concept. The etymology section is in fact the interface between physical and human sciences. The *Etymological Dictionary of Astronomy and Astrophysics* is indeed the first fully fledged etymological dictionary in this field. In particular much effort has been made on the etymology of Persian words dealing also with dialects and other languages of the Iranian branch. The dictionary is careful about the linguistic and terminological aspects of the terms, their morphological structure, and, in a broader scope, the mechanisms that govern a scientific language. The terminological and linguistic analysis of astronomical concepts will be addressed in a separate work.

### 3.2 Multidisciplinarity

Astronomy is tightly related to other branches of knowledge. It even includes other sciences in subfields such as astrobiology, astrochemistry, astrogeology, planetary meteorology, and so on. The dictionary therefore contains a large number of terms in physics, mathematics, geology, meteorology, including philosophy. The hypertext ability enables the reader to move on from a given concept to related ones. Moreover, the dictionary also guides the reader to families of associated concepts in astronomy as well as other fields of knowledge. Owing to new informatics tools resulting from technological advances, we are now able to transcend the partial scope of disciplinary worldviews. Innovative integrated approaches involving synergy from different backgrounds are indispensable for the production and diffusion of knowledge. Initiatives are underway to expand the multidisciplinary aspects of this work. Although more and more specialization for scientific progress is inevitable, inter-, multi-, and transdisciplinary initiatives are necessary to overcome the compartmentation of knowledge (see the dictionary for the definitions of these terms).

### 3.3 Educational and cultural vocations

The dictionary is intent on providing the most recent concepts in astrophysics. It is at the same time careful to explain the previous concepts upon which the new ones are based. This chain of reasoning enables the dictionary to contain the most basic concepts and be self-sufficient. Moreover, such didactic method should make the outcome user-friendly. A new culture based on information is changing the way people learn, work, interact, and live. Education is a key for development. In this new context a great research effort is being made to find technological tools to support the new demands of education. The AstroConcepts semantic tool we have presented in this paper will help the *Etymological Dictionary of Astronomy and Astrophysics* to participate more in this effort.

*Acknowledgements.* We would like to thank Dr. Sébastien Derriere, Centre de Données astronomiques de Strasbourg (CDS), Strasbourg Observatory, for very fruitful discussions and advice. We are also grateful to Dr. Françoise Genova, CDS Director, as well as the INSU Specific Action (ASOV) for their financial support and scientific encouragements.